\documentstyle[pra,aps,twocolumn]{revtex}

\def\QED{\leavevmode\unskip\penalty9999 \hbox{}\nobreak\hfill
     \quad\hbox{\leavevmode  \hbox to.77778em{%
               \hfil\vrule   \vbox to.675em%
               {\hrule width.6em\vfil\hrule}\vrule\hfil}}
     \par\vskip24pt}

\def\ibb #1{\leavevmode\hbox{\kern.3em\vrule
     height 1.5ex depth -.1ex width .4pt\kern-.3em\rm#1}}

\def\Rx {{\ibb R}}

\def\idty{{\rm1\mkern -5.4mu I}}

\title{All multipartite Bell correlation  inequalities for
       two dichotomic observables per site}
 \author{R.~F. Werner\thanks{Electronic Mail: \tt{r.werner@tu-bs.de}}
 {{}\ and\ }M.~M. Wolf\thanks{Electronic Mail: \tt{mm.wolf@tu-bs.de}}
   \\[1ex]
  {\small Institut f{\"u}r Mathematische Physik, TU Braunschweig,}\\
  {\small Mendelssohnstr.3, 38106 Braunschweig, Germany.}}
\date{\today}

\begin{document}
\draft \maketitle

\begin{abstract} We construct a set of  $2^{2^n}$ independent Bell
correlation inequalities for $n$-partite systems with two
dichotomic observables each, which is complete in the sense that
the inequalities are satisfied if and only if the correlations
considered allow a local classical model. All these inequalities
can be summarized in a single, albeit non-linear inequality. We
show that quantum correlations satisfy this condition provided the
state has positive partial transpose with respect to any grouping
of the $n$ systems into two subsystems. We also provide an
efficient algorithm for finding the maximal quantum mechanical
violation of each inequality, and show that the maximum is always
attained for the generalized GHZ state. \end{abstract}


\narrowtext

\section{Introduction}

Entanglement has not only been a key issue in the ongoing debate
about the foundations of quantum mechanics, started by Einstein,
Podolsky and Rosen in 1935 \cite{EPR}. It also  plays a crucial
role in the young field of quantum information theory. Here
entangled states are one of the basic ingredients of quantum
information processing, due to their role as a resource in quantum
key distribution, super dense coding, quantum teleportation and
quantum error correction (cf.\cite{QIPreview}). Although general
structural knowledge about entanglement has improved dramatically
in the last few years, there are still many open problems. For
example, there is still no efficient general method to decide
whether a given state is entangled or not.

The first, and for a long time also the only mathematically sharp
criteria for entanglement were the Bell inequalities \cite{Bell}.
They provided the first possibility to distinguish experimentally
between quantum mechanical predictions and those of local
realistic models. But although Bell inequalities have been  known
for more than thirty years \cite{remark1}, our knowledge about the
precise border between the classical and quantum mechanical
accessible region is still mainly restricted to the simplest
non-trivial cases. Best known is the case of two sites, at each of
which two dichotomic observables are chosen. This is characterized
{\it completely} by the Clauser-Horne-Shimony-Holt (CHSH) version
of Bell's inequalities \cite{CHSH}, in the sense that the
inequalities are satisfied if and only if a local classical model
exists \cite{Fine}. Finding a complete set of linear inequalities
in more complicated situations (more sites, more observables, more
outcomes) turns out to be a very difficult problem in the sense of
computational complexity \cite{Pitovsky}. There is only very
little knowledge about Bell type inequalities beyond the CHSH case
\cite{Mermin,Ardehali,Klyshko,Gisin,problem}. Though numerical
studies yield a large number of inequalities \cite{PS}, for most
of them it is neither known by how much they can be violated in
quantum theory nor is there a general characterization admitting
further investigations.

We were therefore quite surprised ourselves at finding an infinite
sequence of multipartite correlation settings for which we could
develop the theory to be as explicit and complete as in the CHSH
case. Our setting generalizes the CHSH-setting to an arbitrary
number $n$ rather than two different sites, but retains the
constraints of just two observables per site with just two
outcomes each. Thus each of the $n$-participants has the choice of
two observables, each of which can take the values $+1$ or $-1$.
For any choice of observables we then consider the expectation
value of the product of all $n$ signs (a ``full'' correlation
function). A Bell inequality is a linear constraint on the set of
all such expectations, which is valid whenever the correlations
can be obtained from a local classical model, and which cannot be
written as a convex combination of other such constraints.
Examples are the CHSH inequality \cite{CHSH} for $n=2$ and their
generalizations going back to Mermin and others
\cite{Mermin,Ardehali,Klyshko,Gisin} leading to a single
inequality for arbitrary $n$.

We remark that this problem setting could be generalized to
include the expectations not only of the product of all $n$ signs,
but also the products of subsets of signs ( ``restricted''
correlation functions). These data would be sufficient to
reconstruct the full joint probability distributions of signs for
all choices of observables. However, most of the derivations in
this paper do not generalize to this setting, and it is not yet
clear which statements would still be valid (maybe with a
different proof). When we talk of the existence of a classical
model, however, it is understood that such a model would also
determine all restricted correlation functions. The omission of
restricted correlation functions from our setting only means that
we do not consider constraints depending on them.

For this class of multipartite correlations we obtained the
following results:

\begin{itemize}
  \item We construct a set of $2^{2^n}$
Bell inequalities, and show its completeness: the correlations
considered allow a local classical model if and only if all these
inequalities are satisfied (Sec.~\ref{construction}).
  \item The convex set of collections of classical correlation functions
is a $2^n$-dimensional hyper-octahedron, which can be described
alternatively by a single nonlinear inequality (Sec.~\ref{construction}).
  \item We discuss the symmetries connecting different
inequalities and develop a construction scheme, which yields all
$2^{2^n}$ equalities by successive substitutions into the CHSH
inequality (Sec.~\ref{symmetries}).
  \item  We reduce the computation of the maximal quantum violations of
each Bell inequality to a simple variational problem with just one
free variable per site. The maxima are already attained in qubit
systems, more specifically for the $n$-party generalization of the
GHZ state \cite{GHZ}, with a choice of observables depending on
the inequality under consideration (Sec.\ref{quantum}).
  \item We extend this method to a characterization of the convex
body of quantum mechanically attainable correlation functions in
terms of its extreme points, which are also found in the
generalized GHZ state. These results are analogous to those of
Tsirelson \cite{Tsirelson1,Tsirelson2} for the bipartite case.
 \item We characterize the Mermin inequality as that Bell inequality,
which can be violated by the widest margin in quantum theory.
 \item
Sec.\ref{ppt} settles the relationship between the correlation
Bell inequalities and another important entanglement property.  We
show that for states having positive partial transposes with
respect to all their subsystems, all $2^{2^n}$ inequalities are
satisfied, so the correlations in such quantum states can be
explained in the context of a local realistic model. This extends
our earlier result \cite{WernerWolf} for Mermin's inequalities,
and is further supporting evidence for a recent conjecture by
Peres \cite{conjecture}, namely that positivity of partial
transposes should generally imply the existence of local realistic
models. \end{itemize}

In the appendix we will discuss some of the general results
obtained in the sections \ref{construction}, \ref{symmetries},
\ref{quantum} in more detail for the special cases $n=3,4$.

\section{Bell's inequalities and convex geometry}\label{geometry}

Before entering the discussion of Bell inequalities in our special
context it is useful to recall some geometric structures of the
general problem and basic facts concerning the duality of convex
polytopes.

Consider a system decomposed into $n$ independent subsystems.
Suppose further that on each of these subsystems one out of $m$
$v$-valued observables is measured. Thus each of the $m^n$
different experimental setups may lead to $v^n$ different
outcomes, so that the raw experimental data are made up of $(m
v)^n$ probabilities. These numbers form a vector $\xi$ lying in a
space of dimension $(m v)^n$ (minus a few for normalization
constraints). Classically, in a local realistic model, $\xi$ would
be generated by specifying probabilities for each classical {\it
configuration}, i.e., for every assignment of one of the $v$
values to each of the $n m$ observables. Here the ``local''
character of the theory is expressed by the property that the
assignment of a value to an observable at site $k$ does not depend
on the observables chosen at other sites. Every configuration $c$
also represents a possible classical (ideally prepared) state, and
hence a vector $\epsilon_c$ of probabilities. The classical
accessible region, which we will denote by $\Omega$, is thus the
convex hull of $v^{(n m)}$ explicitly known extreme points. Even
though the number of configurations is large, it is finite, hence
$\Omega$ is a {\it polytope}.

Like every compact convex set, $\Omega$ is the intersection of all
half spaces containing it.  A half space is completely
characterized by a linear inequality, so we must look for vectors
$\beta$ such that  $\langle\beta,\xi\rangle\leq1$ for all
$\xi\in\Omega$. Since this property can be checked on the extreme
points $\epsilon_c$ we must look at the convex set
 \begin{equation}\label{Bellset}
 {\cal B}=\{\ \beta \ |\forall c :\langle \beta,
                   \epsilon_c\rangle\leq 1\},
\end{equation}
 also known as the {\it polar} of $\{\epsilon_c\}$. For each
$\beta\in{\cal B}$ the inequality $\langle\beta,\xi\rangle\leq1$
is thus a necessary condition for $\xi\in\Omega$. Moreover, the
Bipolar Theorem \cite{Sc} says that the collection of all these
inequalities is also sufficient.

Luckily, the inequalities are not all independent, since the
inequality for a convex combination $\beta=\sum\lambda_i\beta_i$,
with $\beta_i\in{\cal B}$ already follows from the inequalities
for the $\beta_i$. It therefore suffices to take only the extreme
points of $\cal B$. For a polytope this has a very intuitive
geometrical interpretation: the half spaces determined by extreme
points touch $\Omega$ in a face of maximal dimension. Moreover,
there are only finitely many such maximal faces, which is to say
that $\cal B$ is also a polytope.

The task of finding all Bell inequalities is therefore a special
instance of a standard problem in convex geometry, known as the
{\it hull problem}: given the extreme points $\{\epsilon_c\}$ of a
polytope $\Omega$,  find its maximal faces or, equivalently, the
extreme points of its polar.

The duality between $\cal B$ and $\Omega$ is a generalization of
the duality between regular platonic solids, under which
dodecahedron and the icosahedron, as well as the octahedron and
the cube are polars of each other. A generalized ($d$-dimensional)
octahedron is the unit sphere in a sequence space
$\ell^1(\{1,\ldots,d\})$. Its polar is the unit sphere in the dual
Banach space $\ell^\infty(\{1,\ldots,d\})$, i.e., a
$d$-dimensional hypercube. This is precisely the situation we will
find for the classically accessible region considered in this
paper, where $d=2^n$.

The first to consider the construction of a complete set of Bell
type inequalities as a problem in convex geometry apparently was
M. Froissart \cite{Froissart}. Unfortunately, however, a general
solution for all $(n,m,v)$ is highly unlikely to exist. To find
some extreme points of (\ref{Bellset}) is not so difficult, but
algorithms providing the complete set are likely to run into
serious growth problems already for very small $(n,m,v)$. In fact,
there is a theorem by Pitowsky \cite{Pitovsky} to the effect that,
in a closely related problem, finding all inequalities would also
solve some known hard problems in computational complexity (this
is in fact strongly connected with the notorious $NP=P$ resp.
$NP=coNP$ questions). Pitowsky and Svozil \cite{PS} have recently
performed an extensive numerical search for $n=3$, and published
their result, the coefficients of $53856$  inequalities on their
website. Unfortunately, there is not much generalizable insight
coming out of this kind of work, but it is nice to see what can be
done in this hard numerical problem. For further problems and
partial results in this genre we refer to the problem page
\cite{problem} on our own website.

In what follows we will restrict to the case $(n,m,v)=(n,2,2)$ and
``full'' correlation functions in the sense described in the
introduction.

\section{All Bell correlation inequalities}\label{construction}
\subsection{Basic notation}

Talking about Bell inequalities one usually has in mind
inequalities of the Clauser-Horne-Shimony-Holt form \cite{CHSH}.
These inequalities refer to correlation experiments, in which each
of two parties has the choice of two $\pm 1$ valued observables to
be measured, i.e., $(n,m,v)=(2,2,2)$. Focusing only on full
correlation functions for multi-particle generalizations of such
systems ($(n,m,v)=(n,2,2)$, $n$ fixed arbitrarily) the raw
experimental data are $2^n$ expectation values, each corresponding
to a different experimental setup. Each setup is labeled by the
choice of observables at each site. We parameterize these choices
by binary variables $s_k\in\{0,1\}$ so that $s_k$ indicates the
choice of the $\pm1$-valued observable $A_k(s_k)$ at site $k$.
Each full correlation function is thus the expectation of a
product $\prod_kA_k(s_k)$, and is labeled by a bit string
$s=(s_1,\ldots,s_n)$.

We will consider these expectations as the components $\xi(s)$ of
a vector $\xi$ in a $2^n$-dimensional space. Then any Bell
inequality is of the form
\begin{equation}\label{ineqform}
  \sum_s\beta(s)\xi(s)\leq1\;,
\end{equation}
 where we have normalized the coefficients $\beta$ so that the
maximal classical value is $1$, in accordance with the definition
of polars in Sec.~\ref{geometry}. The linear combination in
Eq.~(\ref{ineqform}) can also be computed under the expectation
value, so that this inequality can be stated as an upper bound on
the expectation of
 \begin{equation}\label{bellpoly}
 B=\sum_s\beta(s)\prod_{k=1}^n  A_k(s_k) .
 \end{equation}
We call such expressions {\it Bell polynomials}. They can be used
directly in the quantum case, where all variables $A_k(s_k)$ are
substituted by operators with $-\idty\leq A_k(s_k)\leq\idty$,
acting in the Hilbert space of the $k$-th site, and the product is
taken as the tensor product. It is often useful to consider these
polynomials rather than the set of coefficients, because often
many coefficients are zero, and we can sometimes simplify a
polynomial algebraically (e.g., by factorization), even though
this may not be apparent from the coefficients.

Two convex sets in the real $2^n$-dimensional vector space are the
subject of our investigation: firstly, the polytope $\Omega$ of
correlation vectors $\xi$ coming from local classical models, and
secondly the set ${\cal Q}\supset\Omega$ of such vectors arising
from quantum models. $\Omega$ will be characterized in terms of
Bell inequalities in this section, ${\cal Q}$ will be considered
in Sec.~\ref{quantum}.

\subsection{Construction and completeness}

In a local classical model every observable $A_k(s_k)$ is a random
variable in its own right, i.e., it is a function of the ``hidden
variable'' which does not depend on the choices $s_\ell$ of
observables at other sites $\ell\neq k$. A model must assign
probabilities to any collection of values for these observables,
i.e., to each {\it classical configuration}. Since the extremal
choices of such probabilities just assign probability $1$ to one
configuration and zero probability to all others, the extreme
points of $\Omega$ are simply labeled by the configurations.

One configuration $c$ is the choice of $c_k(s_k)\in\{-1,1\}$ for
all $k$ and $s_k$´. Clearly, there are $2^{2n}$ such
configurations. The corresponding correlation vector
$\xi\equiv\epsilon_c$ has components
 \begin{equation}\label{epsilonc}
   \epsilon_c(s)=\prod_{k=1}^n c_k(s_k)\;.
\end{equation}
Since we only consider full correlation functions (and not
restricted ones, see the introduction), different classical
configurations may give the same extreme point $\epsilon_c$. For
example, we may choose two different sites, and change the values
of all $c_k(s_k)$ at these sites simultaneously. Then in
Eq.~(\ref{epsilonc}) the sign changes cancel for all $s$. This is
also apparent from the factorization
\begin{equation}\label{verifyreduction}
\epsilon_c(s)=\left(\prod_{k=1}^n c_k(0)\right)\
               \prod_{l=1}^n c_l(0)c_l(s_l),
\end{equation}
in which the first factor is just an $s$-independent sign, and in
the second factor it suffices to choose configurations with
$c_k(0)=1$. Thus we can write $c_k(s_k)=(-1)^{s_kr_k}$ with
$r_k\in\{0,1\}$. Then
\begin{equation}\label{confr}
\epsilon_c(s)=\pm (-1)^{\langle r, s\rangle}, \quad
 \langle r, s\rangle = \sum_{k=1}^n r_ks_k ,
\end{equation}
where the extreme points are now labeled uniquely by the bit
string $r=(r_1,\ldots,r_n)$ and the overall sign. This leaves us
with exactly $2^{n+1}$ extreme points of $\Omega$.

Our task is now to find the extremal linear inequalities $\beta$,
characterizing this set, i.e., the extreme points of ${\cal B}$
from Eq.~(\ref{Bellset}). The bipartite case was indeed completely
analyzed by Fine \cite{Fine}, who showed that there are only two
classes of inequalities: one is trivial in the sense, that it just
requires correlations to be in $[-1,+1]$, and the the second
consists of the CHSH type inequalities, for which the prototype is
$\beta=({1\over2},{1\over2},{1\over 2},-{1\over2})$. A
construction of some Bell type inequalities for arbitrary $n$ was
first proposed by Mermin \cite{Mermin} and further developed by
Ardehali \cite{Ardehali}, Klyshko \cite{Klyshko} and Gisin
\cite{Gisin}.

We will now find {\it all} extremal solutions $\beta$ to the set
of inequalities
\begin{equation}\label{-1beta1}
  -1\leq \sum_s \beta(s)(-1)^{\langle r, s\rangle} \ \leq 1,
\end{equation}
where $r\in\{0,1\}^n$ runs over all bit strings characterizing the
configurations. Suppose that of these $2^n$ inequalities $p<2^n$
fixed ones  are ``tight'' in the sense that the sum takes one of
the extreme values $\pm1$. This will be consistent with a plane
(affine manifold) of vectors $\beta$ of dimension at least
$2^n-p$. We can now construct convex decompositions of $\beta$ in
an open neighborhood of $\beta$ in this plane, since each one of
remaining sums is continuous in $\beta$, and there is a finite
margin before another inequality becomes violated. This
contradicts extremality, so we conclude that the inequality must
be tight for all $r$. Thus we have $2^n$ signs $f(r)\in\{+1,-1\}$
with
\begin{equation}\label{extremalbeq}
\sum_s \beta(s)(-1)^{\langle r, s\rangle} = f(r),\quad
\end{equation}
 Now we can read Equation (\ref{extremalbeq}) as
a Fourier transform with respect to the group of $n$-tuples of
$\{0,1\}$ with addition modulo $2$. Therefore, we easily obtain
the entire set of extremal $\beta$ by applying the inverse
transformation to the set of vectors $f\in\{-1,1\}^{2^n}$:
 \begin{equation}\label{betafourier}
 \beta(s)=2^{-n}\sum_r f(r) (-1)^{\langle r, s\rangle}.
\end{equation}
These are the coefficients of the complete set of $2^{2^n}$
extremal Bell inequalities specifying the range of expectations of
full correlation functions for any local realistic model.

The inequalities constructed in this way have a natural numbering,
defined by the following procedure: For any number between $0$ to
$2^{2^n}-1$, write the binary expansion with ``digits'' $\pm1$ to
get $f$, and perform the inverse Fourier transform
(\ref{betafourier}). From $\beta$ compute the polynomial
(\ref{bellpoly}), which is often the best form of writing the
inequality, because one can apply algebraic simplifications. For
examples of this numbering, see the appendix. The converse
procedure is similar. For example,  the Mathematica package
available from our website \cite{problem} finds that Mermin's
inequality for $n=6$  has the number
$1\;692\;930\;046\;964\;590\;721$.

\subsection{Structure of the classical region}

From the previous section it is clear that the classical region
$\Omega$ is a polytope in $d=2^n$ dimensions with $2d$ extreme
points and $2^d$ maximal faces. This suggests that $\Omega$ should
be a hyper-octahedron, whose polar ${\cal B}$ is a hyper-cube.
Indeed from the parametrization of the inequalities by $d$ values
$f(r)=\pm1$, the latter statement is rather obvious. That $\Omega$
is an octahedron is not so apparent in the coordinates labeled by
$s$ as above. However, we can choose a basis transformation making
this geometric identification of $\Omega$ more obvious. The
necessary transformation is, of course, just the Fourier
transform. With the notation
\begin{equation}\label{xifourier}
\hat{\xi}(r) = 2^{-n}\sum_s (-1)^{\langle r,s\rangle}\xi(s)
\end{equation}
 we can summarize the findings of the previous section by saying
that $\xi\in\Omega$ if and only if
\begin{equation}\label{clmodel1}
\forall f\in\{-1,1\}^{2^n} :\ \sum_r f(r) \hat{\xi}(r)\leq 1.
\end{equation}
 The expression in (\ref{clmodel1}) reaches its maximum with respect
to $f$, if $f(r)$ is just the sign of $\hat{\xi}(r)$. Therefore,
the whole set of $2^{2^n}$ linear inequalities (or the statement
$\xi\in\Omega$) is equivalent to the single non-linear inequality
\begin{equation}\label{clmodel2}
\sum_r|\hat{\xi}(r)|\leq 1.
\end{equation}
Obviously, this nonlinear inequality is nothing but the
characterization of the hyper-octahedron in $2^n$ dimensions as
the unit sphere of the Banach space $\ell^1$.

From this simple characterization of $\Omega$ it might seem that
our problem is essentially trivial. However, the vast symmetry
group of $\Omega$, which includes among other transformations the
set of $(2^n)!$ permutations of the coordinates is misleading,
because these are not really symmetries of the underlying problem
of finding all correlations within a classical model. This is
apparent from the observation that the Bell polynomials associated
with the extreme points may look quite different algebraically.
That is, the $2^n$ dimensions are not really equivalent, but carry
some structure coming from the division of the system into $n$
sites. This is even more obvious when looking at the set of
quantum correlations, which has a much lower symmetry.

Nevertheless, the underlying problem has a large symmetry group,
which will be studied in the next section.

\section{Symmetries and substitutions}\label{symmetries}

Browsing through the complete set of linear correlation
inequalities one quickly gets the feeling that there are many
rather similar ones, and also some inequalities which can be
obtained in a rather trivial way (e.g., as a product) from lower
order ones. In this section we will describe the grouping of the
inequalities into ``essentially different ones'', and also how
they can be obtained by an efficient construction for composing
higher order inequalities from lower order ones. Both ways of
structuring the set of inequalities make sense for more general
cases $(n,m,v)$ (see Section~\ref{geometry}), but for the moment
we only apply them to our restricted class.

\subsection{Symmetry Group}
Some symmetries acting on Bell inequalities are obvious and, in
fact, present in any problem of this type, involving any number of
outcomes and observables. The basic symmetries leading to
equivalent inequalities are:
\begin{enumerate}
\item Changing the labeling of the observables at each site.
\item Changing the names of the outcomes of each observable.
\item Permuting subsystems.
\end{enumerate}
Since we have two observables per site, there are $2^n$ ways of
swapping the labels of observables at each site.  Swapping the
$\pm1$ outcomes of an observable $A_k(s_k)$ at site $k$ results in
a sign in all correlation functions involving this observable. We
have already utilized the fact that swapping both $A_k(0)$ and
$A_k(1)$ only results in an overall sign, so it is enough to
consider sign changes for $A_k(1)$ only. Clearly, there are $2^n$
such sign changes. Expressed in terms of the function $f$ these
transformations amount to
\begin{equation}\label{Weyl}
  f(r)\mapsto (-1)^{\langle s_0, r\rangle}f(r+r_0)\;,
\end{equation}
where $r,s_0,r_0$ all lie in $\{0,1\}^n$, and $r_0$ and $s_0$ are
the parameters describing the sign changes and observable swaps,
respectively. Together with the global sign change and the $n!$
permutations we thus find the group $G$ of symmetry
transformations in our case to have the order
\begin{equation}\label{grouporder}
  |G|= n!\; 2^{2n+1}\;.
\end{equation}
The {\it orbit} of a given inequality is defined as the set of all
the inequalities generated from it by symmetry transformations.
The number of elements in an orbit is $|G|$, divided by the order
of the group of symmetries leaving an element of the orbit
invariant. The number of different orbits is the number of
``essentially different'' inequalities. Obviously,
(\ref{grouporder}) is an upper bound on the number of elements in
each orbit. Since the union of all orbits is the set of all
inequalities, this leads to a lower bound on the number of
essentially different inequalities.

Note that $|G|$ increases much more slowly than $2^{2^n}$, the
total number of extremal inequalities. Therefore, for large $n$
the classification up to symmetry hardly reduces the number of
cases. Explicitly, we find:\par\vskip12pt
\begin{center}\begin{tabular}{c|r|r|r|r} $n$\ \ &\ \ inequalities\ &
$|G|$\ & orbits\ \\ \hline
 2&         16&     64&2\\
 3&        256&    768&5\\
 4&      $65\;536$&  $12\;288$&39\\
 5&\ $ 4\;294\;967\;296$&\ $245\;760$&\ $\geq 17\;476$
\end{tabular}\end{center}
For $n$ up to $4$, the number of orbits was obtained explicitly.
However, for $n\geq5$ the lower bound on the number of orbits
makes it clear that listing all essentially different inequalities
is not going to be useful. More detailed results up to $n=4$ will
be shown in the appendix.

\subsection{Generating new inequalities by substitution}

A simple way of generating inequalities for higher $n$ is to
partition the $n$ sites into two subsets of sizes $n_1$ and
$n_2=n-n_1$ and to take arbitrary Bell polynomials for $n_1$ and
$n_2$ sites, appropriately rename the variables, and to multiply
the two expressions. For example, the polynomial
\begin{equation}\label{3=2x1}
  \frac12( a_1b_1+a_1b_2+a_2b_1-b_1b_2)c_1
\end{equation}
is obtained by multiplying a CHSH polynomial for the first two
sites with the trivial polynomial ``$c_1$'' on the third (note
that for the sake of clarity we have substituted $A_1(0), A_2(1)$
with $a_1, b_2$ etc.). It is clear that this gives an extremal
Bell inequality for three sites.

This procedure can be generalized considerably by noting that the
product operation corresponds to the trivial two site Bell
polynomial ``$a_1b_1$'', but nothing restricts us to using a
trivial expression here. So in general, consider a partition of
the sites into $K$ subsets of sizes $n_k$, $\sum_{k=1}^K n_k=n$.
Then pick an extremal Bell polynomial for $K$ sites, written out
in variables $A_1(0),A_1(1),\ldots,A_K(1)$. Now substitute for
each $A_k(s_k)$ an extremal Bell polynomial for $n_k$ sites. We
claim that the resulting polynomial in $n$ variables is an
extremal Bell polynomial.

Indeed, if we substitute for each of the variables either $+1$ or
$-1$, we will get $A_k(s_k)=\pm1$ for each $k,s_k$, because we
substituted extremal Bell polynomials. But then the same argument
on the level of $K$ sites shows that the value will be $\pm1$.

We will say that a Bell polynomial is {\it elementary}, if it
cannot be obtained by substitution from lower order polynomials.
Obviously, if an inequality is elementary, so is its entire orbit.
Clearly, the CHSH inequality is elementary. Moreover, it is known
that it is a good tool for generating higher order inequalities by
substitution: one of the constructions \cite{Klyshko,Gisin} of the
Mermin's inequalities is based on this idea. But in view of the
rapid increase of the double exponential one might think that
there must be many more elementary inequalities. However, we have
the following result:\vspace{7pt}

{\bf Proposition.\ \it The CHSH-inequality is the only elementary
Bell inequality in the class we consider, i.e., all these
inequalities for $n>2$ can be constructed by successive
substitutions into the CHSH-inequality.}\vspace{7pt}

It is an interesting open problem, whether this statement holds
for other families of Bell inequalities, e.g., the one tabulated
in \cite{PS}.

We start the proof on the level of vectors $f\in\{-1,1\}^{2^n}$
parameterizing an arbitrary extremal Bell inequality for $n$
sites. We decompose the system into a partition of $K=2$ subsets
of size $n-1$, $1$ and rewrite
\begin{equation}\label{fdecomposition}
f(\underbrace{r_1,\ldots,r_{n-1}}_{\widetilde{r}},r_n) =
f(\widetilde{r},0)\delta_{r_n,0} +
f(\widetilde{r},1)\delta_{r_n,1} . \end{equation} The respective
coefficients $\beta(s)$ of the $n$-site inequality are then
obtained via Fourier transformation according to Eq.
(\ref{betafourier}), and we get
\begin{eqnarray}\label{betatolower} &\beta(s)&\ =\ 2^{-n}\sum_{r}
f(r)(-1)^{\langle r, s \rangle}\nonumber\\ &=& \frac12
\beta_0(\widetilde{s})\sum_{r_n}(-1)^{s_n r_n}\delta_{r_n,0} +
\frac12 \beta_1(\widetilde{s})\sum_{r_n}(-1)^{s_n
r_n}\delta_{r_n,1}\nonumber\\ &=& \frac12 \Big[
\beta_0(\widetilde{s})+(-1)^{s_n}\beta_1(\widetilde{s})\Big] ,
\end{eqnarray} where $\beta_k(\widetilde{s})$ are coefficients for
extremal Bell inequalities for $n-1$ sites. If we now add the
respective observables $A_k(s_k)$ and write out the corresponding
Bell polynomial \begin{eqnarray}\label{CHSHsubstitute} B &=&
\sum_s \beta(s) \prod_{k=1}^n A_k(s_k)\nonumber\\ &=& \frac12 B_0
\Big[A_n(0)+A_n(1)\Big]+\frac12 B_1\Big[A_n(0)-A_n(1)\Big],
\end{eqnarray} we immediately see, that this is just a CHSH
polynomial, where the observables of one site have been
substituted by Bell polynomials $B_0$ and $B_1$ for $n-1$ sites.

\section{Quantum violations}\label{quantum}

Provided with a huge number of Bell type inequalities we now go
beyond the classical accessible region. The first question to
arise is of course whether or not and to what extent quantum
systems can violate these inequalities. To answer this question we
will first provide an effective variational method for computing
the maximal quantum violations and show, that they are bounded by
those obtained for Mermin's inequalities. In the following two
subsections we will then briefly discuss the structure of the
underlying quantum domain, and prove that the generalized GHZ
state maximally violates any of the correlation inequalities.

\subsection{Obtaining the maximal violations}\label{qmax}

In order to compute the maximal quantum violation of any
correlation inequality we have to vary over one density operator
$\rho$ on a tensor product of $n$ factors, and two operators in
each factor. Assuming all tensor factors to have dimension $d$,
this means $d^{2n}$ parameters for the density operator and
$2nd^2$ for the observables. Hence the numerical solution of this
variational problem is not feasible, except for the most trivial
cases (and even impossible, because $d$ is, in principle, a free
parameter).  Fortunately, however, it turns out that computing the
overall maximum is much easier than computing the maximal
violation for a fixed state:  we will reduce the computation to a
variational formula in just $n$ variables.

First we have to recall some basic notions. In quantum mechanics
expectations of $\pm 1$-valued observables are described by
Hermitian operators $A_k(s_k)$ with spectrum in $[-1,+1]$. Since
we are only interested in maximal correlations, we may as well
take the observables extremal in the convex set of Hermitian
operators with $-{\idty}\leq A\leq\idty$, i.e., we may assume the
observables to be unitary and thus $A^2=\idty$.

The general form of a Bell inequality for an $n$-partite quantum
system, which is characterized by a density operator $\rho$, is
then
\begin{equation}\label{QBell}
{\rm tr}(\rho B)\mathpunct:={\rm tr}\Bigl[ \rho
\sum_{s}\beta(s)\bigotimes_{k=1}^n A_k(s_k)\Bigr]\leq 1 ,
\end{equation}
where we will refer to $B$ as the {\it Bell operator}, which is
just the quantum counterpart of the Bell polynomial defined in
Eq.(\ref{bellpoly}). Of course, every expectation value
(\ref{QBell}) larger than $1$ is called a violation of Bell's
inequality.

In order to derive the maximal quantum violation, which is nothing
but the operator norm of the Bell operator, we first define
another operator $C$ by
\begin{equation}\label{C}
 C:=B\bigotimes_{k=1}^nA_k(0)
  =\sum_s\beta(s)\bigotimes_{k=1}^nC_k^{s_k}\;,
\end{equation}
 where we have set $C_k=A_k(1)A_k(0)$, and $C_k^0=\idty$. Since the
$C_k$ are commuting unitary operators, all summands of $C$ can be
diagonalized simultaneously, and the eigenvectors of $C$ are
tensor products of eigenvectors of the $C_k$.  Every eigenvalue
$\gamma$ of $C$ is therefore of the form
\begin{equation}\label{evgamma}
  \gamma = \sum_s\beta(s)\prod_{k=1}^n\gamma_k^{s_k},
\end{equation}
 where $\gamma_k$ is an eigenvalue of $C_k$. It is clear from the above
remarks that $C$ commutes with its adjoint, so $\|C\|$ is just the
modulus of the largest eigenvalue.  Now we utilize the fact that
$\|B^*B\|=\|C^*C\|$, i.e., $\|B\|=\|C\|$ and obtain
 \begin{equation}\label{normB}
  \|B\| = \sup_{\{\gamma_k\}} \Big|\sum_s\beta(s)\prod_{k=1}^n
             \gamma_k^{s_k}\Big| ,
\end{equation}
 where each $\gamma_k$ runs over the eigenvalues of $C_k=A_k(1)A_k(0)$.
This formula allows us to compute the largest expectation ${\rm
tr}(\rho B)$ for fixed real coefficients $\beta$ (coming from a
Bell inequality or not) and a fixed choice of observables
$A_k(s_k)$, but with $\rho$ chosen without further constraints to
maximize the expectation.

What we are now interested in is the maximum also with respect to
the $A_k(s_k)$. Since formula (\ref{normB}) depends only on the
eigenvalues $\gamma_k$ this will be given by the same expression,
but with $\gamma_k$ running not just over the eigenvalues of a
particular operator $C_k$ but over all $\gamma_k$ which can be
eigenvalues of products of unitary and hermitian operators. Since
such a product is again unitary, we have $|\gamma_k|=1$. Moreover,
as is easily seen in $2\times2$ examples, this is the only
constraint in any Hilbert space dimension (see also
Sec.~\ref{GHZsec}). Hence for any choice of real coefficients
$\beta(s)$ and observables $-\idty\leq A_k(s_k)\leq\idty$, we
have, as the best possible bound:
 \begin{equation}\label{maxall}
   {\rm tr}\Bigl[ \rho \sum_{s}\beta(s)\bigotimes_{k=1}^n
                   A_k(s_k)\Bigr]
     \leq \sup_{\{\gamma_k\}} \Big|\sum_s\beta(s)\prod_{k=1}^n
             \gamma_k^{s_k}\Big| ,
\end{equation}
where the supremum runs over all $\{\gamma_1,\ldots,\gamma_n\}$
with $|\gamma_k|=1$. Moreover, the bound does not change with
Hilbert space dimension, as long as all factors are non-trivial.

A more detailed discussion of quantum violations utilizing
Eq.~(\ref{normB}) for the special cases $n=3,4$ can be found in
the appendix.

\subsection{Mermin's inequalities and the overall maximum}

Asking for the overall maximal quantum violation we may
additionally vary over the set of inequalities. In utilizing the
result obtained in the previous section, we are able to express
the norm of a Bell operator in terms of lower order Bell
operators. Moreover it suffices to consider qubit systems and we
may therefore set $A_k(s_k)=\vec{a}_k(s_k)\vec{\sigma}$, where
$\vec{\sigma}$ is the vector of Pauli matrices and
$\vec{a}_k(s_k)$ is a normalized vector in $\Rx^3$.

Squaring Eq.(\ref{CHSHsubstitute}) this leads to
\begin{eqnarray}\label{overallviolation}
B^2&=&
\frac{B_0^2}2\otimes\big[1+\vec{a}_n(0)\vec{a}_n(1)\big]+\frac{
B_1^2}2\otimes\big[1-\vec{a}_n(0)\vec{a}_n(1)\big]\nonumber\\
&+&[B_0,B_1]\otimes\frac{i}2
[\vec{a}_n(0)\times\vec{a}_n(1)]\vec{\sigma} .
\end{eqnarray}
Without loss of generality we now assume that $\|B_0\|\leq\|B_1\|$
and estimate by induction
\begin{equation}\label{violbound}
\|B\|^2=\|B^2\|\leq 2\|B_1\|^2\leq 2^{n-1}.
\end{equation}

This bound is indeed saturated by the set of inequalities going
back to Mermin \cite{Mermin,Ardehali,Klyshko,Gisin}, which thus
provide the overall maximal quantum violation. In fact, we will
show, that the converse is also true, so that we have the
following \vspace{7pt}

{\bf Proposition.\ \it The orbit corresponding to Mermin's
inequality is the only one for which the maximal violation
$2^{\frac{n-1}2}$ is attained. }\vspace{7pt}

Before we continue proving the claimed uniqueness, we emphasize,
however, that this does in general not imply, that for a fixed
quantum state Mermin's inequality is more strongly violated than
any other.

We begin our proof with noting that the maximal norm of the Bell
operator in Eq.(\ref{overallviolation}) requires orthogonality of
the observables, such that the respective phases in
Eq.(\ref{normB}) have to be $\pm i$. Without loss of generality we
can thereby restrict to the case $+ i$ since the remaining sings
just correspond to a transformation between two inequalities of
the same orbit according to (\ref{Weyl}). Hence, Eq.(\ref{normB})
leads to
\begin{eqnarray}\label{Merminup}
\|B_{max}\| &=& 2^{-n}\Big|\sum_{r,s}f_{max}(r)\prod_{k=1}^n
(-1)^{s_k r_k}i^{s_k}\Big| \nonumber\\&=& 2^{-n}\Big|\sum_r
f_{max}(r) \prod_{k=1}^n \big[1+i (-1)^{r_k}\big]\Big|\nonumber\\
&=& 2^{-\frac{n}{2}}\Big|\sum_r f_{max}(r)\
\prod_{k=1}^n\big[e^{i\frac{\pi}{4}(1-2r_k)}\big]\Big|\nonumber\\
&=&  2^{-\frac{n}{2}}\Big|\sum_r f_{max}(r)\ (-i)^{\sum_k
r_k}\Big|.
\end{eqnarray}
If we now want $B_{max}$ to saturate the bound in
(\ref{violbound}), then following Eq.(\ref{Merminup}) we are left
with four possible choice for $f_{max}$, like $f_{max}(r)= 1$ for
$(-i)^{\sum_k r_k}= 1, i$ and $f_{max}(r)= - 1$ otherwise. Since
these four inequalities again belong to the same orbit, the
correlation inequality leading to the overall maximal quantum
violation is indeed uniquely determined (up to equivalence
transformations within one orbit).

\subsection{Structure of the quantum domain}\label{qdomain}

In the same manner as we did for the classical case we may ask for
the structure of the region in the space of correlations, which is
accessible within the framework of quantum mechanics. One of the
first to investigate this question in more detail apparently was
Tsirelson \cite{Tsirelson1,Tsirelson2}, while studying quantum
generalizations of Bell's inequalities.

Let us begin with defining the quantum counterpart of the
classical accessible region $\Omega$, introduced in Sec.
\ref{geometry}: \begin{equation}\label{Q} {\cal Q}:=\bigg\{\;\xi\;
\Big|\; \xi_s={\rm tr}\Big[\rho\bigotimes_{k=1}^n
A_k(s_k)\Big]\bigg\}\subset \Rx^{2^n}, \end{equation} where
$\{A_k\}$ are suitable observables and $\rho$ is a quantum state
in arbitrary dimension. The structure of $\cal Q$ is much more
complicated than that of $\Omega\subset\cal Q$. In particular, it
is not a polytope. Nevertheless, we can explicitly parameterize
its extreme points. For the sake of completeness we will first
prove convexity of $\cal Q$, although this follows closely the
work of Tsirelson \cite{Tsirelson2}.

Consider a convex combination of vectors in $\cal Q$
\begin{equation}\label{Qconvex}
\sum_\alpha \lambda^{(\alpha)}\xi^{(\alpha)} , \quad
\xi^{(\alpha)}\in\cal Q
\end{equation}
and an associated Hilbert space
\begin{equation}\label{Hilbert}
{\cal H}=\bigotimes_{k=1}^n{\cal H}_k=\bigotimes_k\bigoplus_\alpha
{\cal
H}_k^{(\alpha)}\cong\bigoplus_{\alpha_1\ldots\alpha_n}\!\!\bigotimes_k
{\cal H}_k^{(\alpha_k)}.
\end{equation}
Then with $\rho=\bigoplus_\alpha
\lambda^{(\alpha)}\rho^{(\alpha)}$, which is a density operator
acting on the ``diagonal subspace''
\begin{equation}\bigoplus_\alpha\bigotimes_k{\cal
H}_k^{(\alpha)}\subset\cal H ,
\end{equation}
and $A_k(s_k)=\bigoplus_\alpha A_k^{(\alpha)}(s_k)$ we are given a
state and observables such, that the convex combination in
(\ref{Qconvex}) is indeed a proper element of $\cal Q$. Hence,
$\cal Q$ is convex.

Now let us return to the result obtained in the previous
subsection. Following Eq. (\ref{normB}) we can write the maximal
quantum violation of an arbitrary inequality $\beta$ as
\begin{eqnarray}\label{Qvio2}
&&\sup_{\gamma_0\ldots\gamma_n}\sum_s\beta(s){\Re\!e}
\Big(\gamma_0\prod_{k=1}^n\gamma_k^{s_k}\Big)\\ &=&
\sup_{\varphi_0\ldots\varphi_n}\sum_s\beta(s)\xi_s(\varphi_0,\ldots,\varphi_n),
\end{eqnarray}
where we have set
$\xi_s(\varphi_0,\ldots,\varphi_n)=\cos(\varphi_0+\sum_k\varphi_ks_k)$.
Now by the Bipolar Theorem \cite{Sc} the convex set $\cal Q$ is
just given by the convex hull of these vectors:
\begin{equation}\label{Qhull}
{\cal Q}=\mbox{co}\big\{\xi(\varphi_0,\ldots,\varphi_n)\big\}.
\end{equation}

\subsection{Generalized GHZ states}\label{GHZsec}

It is a well known fact, that the generalized GHZ state defined by
\begin{equation}\label{GHZ}
|\Psi_{GHZ}\rangle = \frac1{\sqrt{2}}\big(|00\ldots 0\rangle +
|11\ldots 1\rangle\big)
\end{equation}
maximally violates Mermin's inequalities \cite{Klyshko}.
Astonishingly this is also true for any other of the $2^{2^n}$
correlation inequalities:\vspace{7pt}

{\bf Proposition.\ \it Any extreme point of the convex set of
quantum correlation functions as defined in Eq.(\ref{Q}) is
already obtained for the generalized GHZ state. In particular,
this implies that GHZ states maximally violate any of the
presented correlation inequalities. }\vspace{7pt}

We have to show, that for any set of angles
$\{\varphi_0,\ldots,\varphi_n\}$ there are suitable observables
such, that
\begin{equation}\label{toshow}
 \langle
\Psi_{GHZ}|\bigotimes_{k=1}^n
A_k(s_k)|\Psi_{GHZ}\rangle=\cos(\varphi_0+\sum_k\varphi_ks_k) .
\end{equation}
Therefore we choose observables
$A_k(s_k)=\vec{a}_k(s_k)\vec{\sigma}$ with
\begin{eqnarray}\label{xyobservables}
\vec{a}_k(0) &=& \big(\cos\alpha ,\ \sin\alpha ,\ 0
\big)\nonumber\\ \vec{a}_k(1) &=& \big(\cos(\varphi_k + \alpha),\
\sin(\varphi_k + \alpha),\ 0\big) .
\end{eqnarray}
These observables simply swap the basis vectors providing them
with an additional phase factor, i.e.,
\begin{equation}\label{01action}
\vec{a}_k(s_k)\vec{\sigma}\ |j\rangle =
\exp\big[i(-1)^j(\alpha+\varphi_k s_k)\big]\ |j\oplus 1\rangle,
\end{equation}
where $j=0,1$ and $\oplus$ means addition modulo 2. Hence, for the
left hand side of Eq.(\ref{toshow}) two terms occur, which are
just complex conjugates of each other, and we get
\begin{equation}\label{xyghz}
\langle\Psi_{GHZ}|\bigotimes_{k=1}^n
A_k(s_k)|\Psi_{GHZ}\rangle={\Re\! e}\Big\{e^{i\alpha
n}e^{i\sum_k\varphi_k s_k}\Big\},
\end{equation}
so that it just remains to set $\alpha=\varphi_0 / n$.

\section{States with positive partial transposes}\label{ppt}

The violation of one of the inequalities, which can be derived
from Eq.(\ref{betafourier}), is a rather physical entanglement
criterion, since we can at least in principal decide it
experimentally by measuring the respective correlations. However,
the difficulty in doing so is the choice of the observables, and
optimizing them for a fixed state leads in general to a very high
dimensional variational problem. An entanglement criterion, which
is in contrast easy to compute, is the {\it partial transpose}
proposed by Peres in \cite{Peres}.

Before we settle the relationship between these two entanglement
criteria, we will briefly recall some basic notions.

The {\it partial transpose} of an operator on a twofold tensor
product of Hilbert spaces ${\cal H}_1\otimes{\cal H}_2$ is defined
by
\begin{equation}\label{PTdef}
\Bigl(\sum_j C_j\otimes D_j\Bigr)^{T_1}=\sum_j C_j^T\otimes D_j,
\end{equation}
where $C_j^T$ on the right hand side is the ordinary transposition
of matrices with respect to a fixed basis. The generalization of
this definition to an $n$-fold tensor product is straight forward,
and we will denote the transposition of all sites belonging to a
set $\tau\subset\{1,\ldots,n\}$ by the superscript $T_\tau$.

Recall further, that a state is called {\it separable} or {\it
classically correlated}, if it can be written as a convex
combination of tensor product states -- otherwise it is called
{\it entangled}. A necessary condition for separability, which
also turned out to be sufficient in the case of two qubits
\cite{2x2}, but not in general (cf.\cite{BE}), is the positivity
of all partial transposes with respect to all subsystems.
Moreover, there is a conjecture by Peres \cite{conjecture}, that
this might even imply the existence of a local realistic model. In
\cite{WernerWolf} we showed, that the set of inequalities going
back to Mermin \cite{Mermin} is indeed fulfilled for states
satisfying this ``ppt''-condition. In the following we will show
that this implication is not due to a special property of these
inequalities, but holds for any Bell type inequality in
(\ref{QBell}), as long as we consider expectations of full
$n$-site correlations. This leads to the main result of this
section:\vspace{7pt}

{\bf Proposition } {\it Consider an $n$-partite quantum system,
where each of the parties has the choice of two dichotomic
observables to be measured. Assume further, that the partial
transposes with respect to all subsystems of the corresponding
density operator are again positive semi-definite operators. Then
all the $2^n$ correlations can be described in the context of a
local realistic model. }\vspace{7pt}

In particular, this implies, that if a state is biseparable with
respect to all partitions, all the inequalities are satisfied even
if there exists no convex decomposition into $n$-fold product
states.

In order to prove this proposition and to derive an upper bound
for the expectation of the Bell operator, we first apply the
variance inequality to $\rho^{T_\tau}$ and $B^{T_\tau}$:
\begin{eqnarray}
 ({\rm tr}\rho B)^2
     &=&\bigl({\rm tr}\rho^{T_\tau} B^{T_\tau}\bigr)^2
      \leq{\rm tr}\bigl[\rho^{T_\tau} (B^{T_\tau})^2\bigr]
 \nonumber\\
     &\leq&{\rm tr}\Bigl\{\rho
     \bigl[(B^{T_\tau})^2\bigr]^{T_\tau}\Bigr\}.
\label{var}
\end{eqnarray}
Since we suppose that $\rho^{T_\tau}\geq 0\ \forall\tau$ this
holds for any partial transposition, and we may take the average
with respect to all subsets $\tau$, and have therefore to estimate
the expectation of the operator
\begin{eqnarray}\label{inanotina}
&&{1\over2^n}\sum_\tau
    \sum_{s,s'}\beta(s)\beta(s')\bigotimes_{k\in\tau}A_k(s_k)A_k(s_k')
    \bigotimes_{k\notin\tau}A_k(s_k')A_k(s_k)\nonumber\\
&&=\sum_{s,s'}\beta(s)\beta(s')\bigotimes_{k=1}^n {1\over 2}
    \big\{A_k(s_k),A_k(s_k')\big\}_+,
 \end{eqnarray}
where $\{\cdot,\cdot\}_+$ denotes the anti-commutator. Note that
in the first line of Eq.(\ref{inanotina}) we have rearranged the
tensor product and made use of
\begin{equation}\label{TTT}
\big[A_k(s_k')^T A_k(s_k)^T\big]^T = A_k(s_k) A_k(s_k').
\end{equation}
Since $A^2=\idty$ and $s_k,s_k'\in\{0,1\}$ only two different
operators can arise in every tensor factor in (\ref{inanotina}):
either ${1\over 2}\{A_k(0),A_k(1)\}_+$ or the identity operator.
These two obviously commute, and we can therefore simultaneously
diagonalize all the summands. What remains to do, is to
substantiate our intuition that ``if everything commutes, then we
are in the classical regime''. For this purpose note, that
eigenvalues of the operator (\ref{inanotina}) are of the form
\begin{equation}\label{EV}
\sum_{s,s'}\beta(s)\beta(s')\prod_{k=1}^n \Bigl\{
  \begin{array}{crccc}
    \chi_k&, s_k&\neq&s_k'& \\
    1&, s_k&=&s_k'& \
  \end{array}\Bigr\} ,
\end{equation}
for suitable $-1\leq\chi_k\leq 1$. But since we can always find
classical observables $\cal C$ with correlations
\begin{equation}\label{cc}
\langle {\cal C}_k(0){\cal C}_k(1)\rangle = \chi_k ,
\end{equation}
we are able to construct a system, which is classical in the sense
that it may be described in the context of classical probability
theory, such, that (\ref{EV}) is the expectation of the square of
the respective Bell polynomial. However, due to the defining
properties of the Bell inequalities, this is indeed bounded by
unity, which proves our claim, that all the considered Bell
inequalities are satisfied for states having positive partial
transposes with respect to all their subsystems.

\section{Conclusion}

We provided two approaches for constructing the entire set of
multipartite correlation Bell inequalities for two dichotomic
observables per site: the Fourier transformation of a $2^n$-digit
binary number and nesting CHSH inequalities. This set of
inequalities led us to a single non-linear inequality, which
detects the existence of a local classical model with respect to
the considered correlations. We were able to simplify the
variational problem of obtaining the maximal quantum violation of
the linear correlation inequalities, in particular showing, that
these are attained for generalized GHZ states, and proved, that
``ppt states'' satisfy all these $2^{2^n}$ inequalities.

One crucial assumptions was, that each site has the only choice of
two dichotomic observables to be measured. Permitting more
observables per site, more outcomes per observable or even the
choice of `` not measuring'', i.e., including restricted
correlation functions, would lead to non-commuting terms, and most
of the arguments would fail. So this is obviously a starting point
for further investigations. In particular, one may think of
applying the mechanism of substitution (Sec.\ref{symmetries}) in
order to derive new classes of Bell inequalities.

Another open question concerns the hierarchy of the inequalities
with respect to their quantum violations. That is, if a given
inequality is violated for a fixed quantum state, is there a set
of inequivalent inequalities, which have to be violated as well?

Finally, we want to mention, that there is recent work by Scarani
and Gisin \cite{ScaraniGisin} pointing out, that there might be a
close relation between the quantum violation of multipartite Bell
inequalities and the security of $n$-partner quantum
communication.

\section*{Acknowledgement}
Funding by the European Union project EQUIP (contract
IST-1999-11053) and financial support from the DFG (Bonn) is
gratefully acknowledged.

\appendix
\section*{}
Recently more and more attention has turned to tri- and
four-partite states, especially to symmetric states as
laboratories for multipartite entanglement (cf.
\cite{Duer,EggelingWerner,KalleChef}). Therefore we will provide
the complete set of Bell inequalities for these cases in a more
explicit form and additionally give the maximal quantum
violations, which we have numerically \cite{numerics} obtained
utilizing the method presented in Sec.\ref{quantum}.
\subsection{Inequalities for three sites}

For $n=3$ Eq.(\ref{betafourier}) leads to the five essentially
different Bell polynomials (for the sake of legibility we again
substitute $A_1(0),A_2(1)$ with $a_1, b_2$ etc.):
 \begin{eqnarray}
&&a_1b_1c_1\label{a1}\\ && {1\over4}\sum_{k,l,m}a_kb_lc_m -
a_1b_1c_1\label{a2}\\ &&
\frac12\bigl[a_1(b_1+b_2)+a_2(b_1-b_2)\bigr]c_1 \label{a3}\\
&&{1\over2}\bigl[a_1b_1(c_1+c_2)-a_2b_2(c_1-c_2)\bigr]\label{a4}\\
&&{1\over2}\bigl(a_1b_1c_2+a_1b_2c_1+a_2b_1c_1-a_2b_2c_2\bigr)\label{a5}
 \end{eqnarray}
(\ref{a1}) and (\ref{a3}) are just trivial extensions of lower
order inequalities, and (\ref{a5}) belongs to the set developed by
Mermin \cite{Mermin}. The maximal quantum violations, the number
of the first inequality of each of the 5 orbits, and the sizes of
the respective orbits are stated in the following table:
\begin{center}
\begin{tabular}{|r|r|c| r} \cline{1-3}ineq.&\ $|$orbit$|$&
qm.viol.&\\ \cline{1-3} 0&16&1&\ \ (\ref{a1}) \\ 1&128&
$5/3$&(\ref{a2})\\ 3&48& $\sqrt{2}$&(\ref{a3})\\ 6&48&
$\sqrt{2}$&(\ref{a4})\\ 23&16&2&(\ref{a5})
\\ \cline{1-3}
\end{tabular}
\end{center}

\subsection{Inequalities for four sites}

For $n=4$ we just give the number of the first inequality of each
of the 39 orbits, its size, and the respective maximal quantum
violations. The index $p$ labels orbits including an element which
is invariant under permutations of the subsystems, and $f$
indicates factorizing Bell polynomials (like (\ref{a1}) and
(\ref{a3}) for tripartite systems):

\begin{center}\begin{tabular}{
| l|r|c|c|l|r|c|}\cline{1-3}\cline{5-7} ineq.&\ $|$orbit$|$&
qm.viol.&\hspace{12pt} & 283&\ \  6144&\ 2.078\  \
\\ \cline{1-3}$0_{p,f}$ & 32&1&& 286&1536&2.078\\
 $1_p$ &512&1.843&&287&1536&2.326\\
 $3_f$&1024&5/3&&300&3072&2\\
 6&1536&5/3&&301&6144&5/3\\
 7&3072&1.932&&303&3072&1.819\\
 $15_f$&192&$\sqrt{2}$&&317&3072&2
 \\22&2048&1.932&&318&1536&2\\
23&1024&$\sqrt{5}$&&319&2048&2.139\\ 24&1024&2&&360&1024&2.326\\
25&6144&$\sqrt{3}$&&363&1536&$\sqrt{3}$\\
27&3072&$\sqrt{3}$&&367&1536&$\sqrt{3}$\\
30&3072&$\sqrt{3}$&&$383_p$&256 & 2\\
$60_f$&384&$\sqrt{2}$&&$831_f$&128&2\\ 105&128&
$\sqrt{2}$&&$854_f$&96&2\\
 $278_p$ &256&$\sqrt{5}$&&857&384& $\sqrt{2}$\\
$279_p$&512&2.556&&874&384&2\\ 280&3072&2.139&&1632&96&
$\sqrt{2}$\\ 281&1536&1.819&&1647&192&2\\ 282&3072&1.819&&
$6014_p$ &32& $2\sqrt{2}$\\ \cline{1-3}\cline{5-7}
\end{tabular}\end{center}

The number of the inequality representing the orbit of Mermin's
inequality is 6014.


\begin{references}
\bibitem{remark1} Of course mathematicians started to investigate the possible range
of correlations in the form of inequalities long before physicists
payed attention to it due to the work of Bell \cite{Bell}. For a
better review cf. Ref. \cite{Pitovsky}.
\bibitem{numerics} Especially for the case $n=4$ we
emphasize that the maximal violations in the table are  results of a numerical
maximization. Although they seem very stable with respect to variation of
the initial conditions, we have no proof that these are the absolute maxima.
\end{references}
\end{document}